\documentclass{blois}

\def\be{\begin{equation}}
\def\ee{\end{equation}}
\def\bea{\begin{eqnarray}}
\def\eea{\end{eqnarray}}

\newcommand{\Photo}{}

\begin{document}
\vspace*{4cm}
\title{$B$-PHYSICS ANOMALIES: EFT ANALYSES AND SIMPLIFIED MODELS}

\author{ L. ALLWICHER }

\address{Physik-Institut, Universit\"at Z\"urich, Winterthurerstrasse 190 \\ CH-8057 Zurich, Switzerland}

\maketitle\abstracts{We discuss some theoretical aspects related to the $B$-anomalies, both for the neutral current and the charged current anomalies. A possible combined explanation within an EFT framework as well as an explicit simplified model featuring the $U_1$ leptoquark are discussed. The model gives rise to predictions in other sectors that are experimentally accessible within the next few years.}

\section{Introduction}

In the Standard Model (SM), Lepton Flavour Universality (LFU) is an almost exact property. This is a consequence of the gauge interactions being flavour-blind, i.e. all generations interact with the $SU(3) \times SU(2) \times U(1)$ gauge fields in the same way.
The only breaking of LFU occurs through the Higgs Yukawa interactions with the fermions, responsible for the different masses.
These are governed by the Yukawa couplings, the biggest of which among the leptons is $y_\tau \sim 10^{-2} \ll 1$.
Hence the expectation is, in the SM, that all processes involving electrons, muons and taus will be the same (up to kinematical factors).
Tests of LFU can therefore be a powerful probe of physics Beyond the Standard Model (BSM).
In particular, it is interesting to look at New Physics (NP) trying to explain also the flavour structure of the SM, i.e. something that is LFU violating by definition.

In recent years, a set of measurements has been challenging LFU in transitions involving $B$ mesons. These generally go under the name of $B$-anomalies, and can be further divided into two main categories, characterised by the underlying quark transition \cite{Cornella:2021sby}.
\begin{itemize}
\item \textit{Neutral current} ($b \to s \ell \ell$) \textit{anomalies}. The relevant observables are the LFU ratios\footnote{In the SM these are all expected to be equal to one \cite{Bordone:2016gaq}.}
  \be
    R_{X_s} = \frac{\mathcal{B}(B \to X_s \mu^+ \mu^-)}{\mathcal{B}(B \to X_s e^+ e^-)} \,,
  \ee
  where $X_s$ indicates a generic strange meson,
  as well as the branching fractions $\mathcal{B}(B_s \to \mu^+ \mu^-)$ and $\mathcal{B}(B \to X_s \mu^+ \mu^-)$, and the angular distribution in the decay $B \to K^* \mu^+ \mu^-$. Among these, it is customary to make the distinction between so-called ``clean'' observables, which are not affected by hadronic uncertainties in the theoretical prediction, and the other observables. The LFU ratios and $\mathcal{B}(B_s \to \mu^+ \mu^-)$ fall into the first category.
  The most recent experimental results have been presented at this conference by John Smeaton. It is however worth mentioning that in 2021 for the first time a $3.1\sigma$ deviation from the SM has been observed in $R_K$ ($R_K^{\mbox{\tiny exp}} = 0.864^{+0.044}_{-0.041}$ \cite{LHCb:2021trn}).
  All other $b \to s \ell \ell$ observables seem to go in the same direction, indicating an overall deficit of muons with respect to the SM expectation.
\item \textit{Charged current} ($b \to c \ell \nu$) \textit{anomalies}. Here the observables are the two LFU ratios
  \be
    R_{D^{(*)}} = \frac{\mathcal{B}(B \to D^{(*)} \tau \nu)}{\mathcal{B}(B \to D^{(*)} \ell \nu)} \,,
  \ee
  with $\ell = e, \mu$, and the measurements signal an excess of taus with respect to muons and electrons. 
\end{itemize}

\section{EFT for the $B$-anomalies: a bottom-up approach}

Under the hypothesis of some heavy NP responsible for these deviations, at energies around the $B$ meson mass the effect is best best parametrised by means of an Effective Field Theory (EFT), and in particular by four-fermion semileptonic interactions.

\begin{figure}
  \centering
  \includegraphics[width=0.4\textwidth]{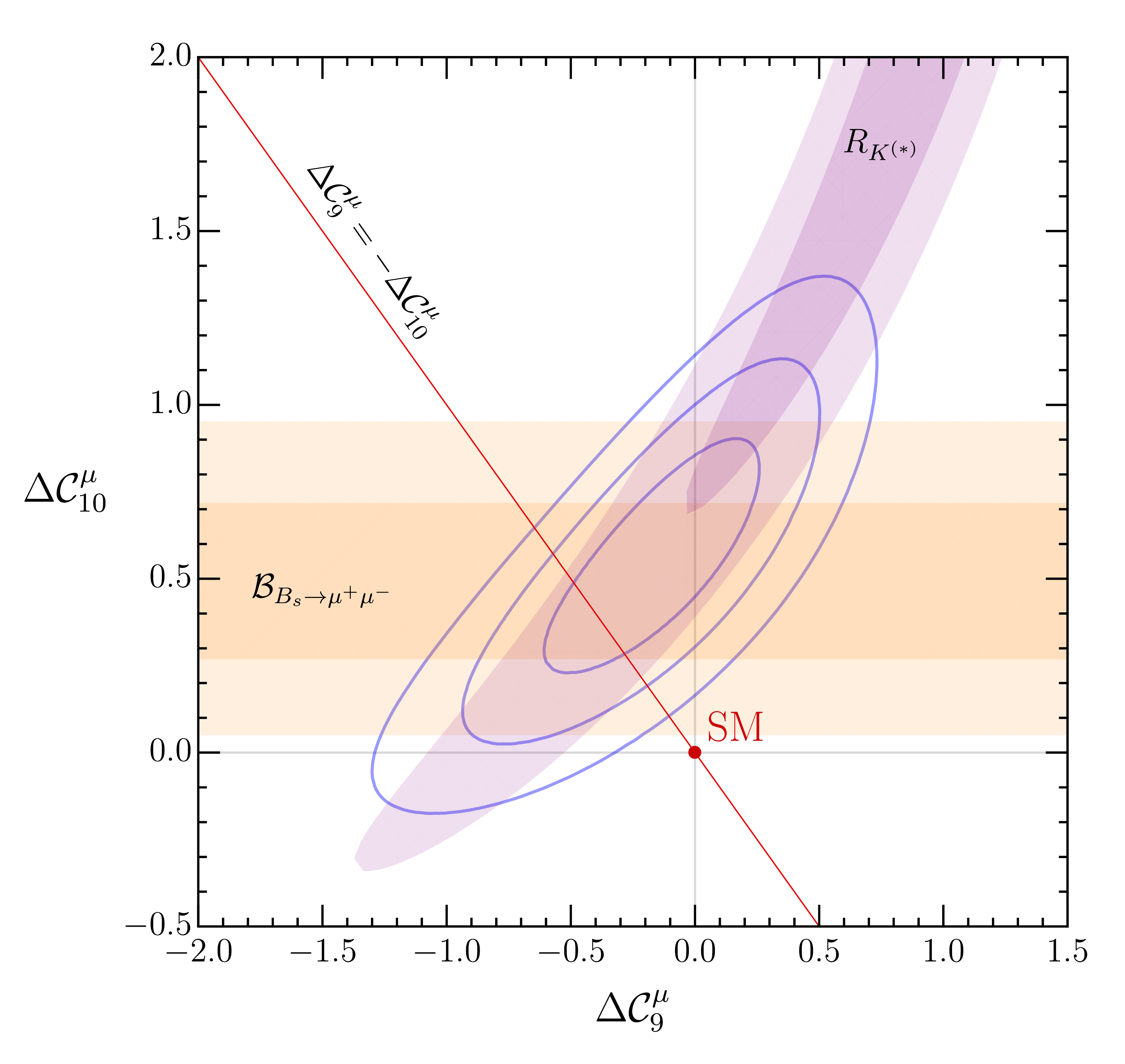}
  \includegraphics[width=0.4\textwidth]{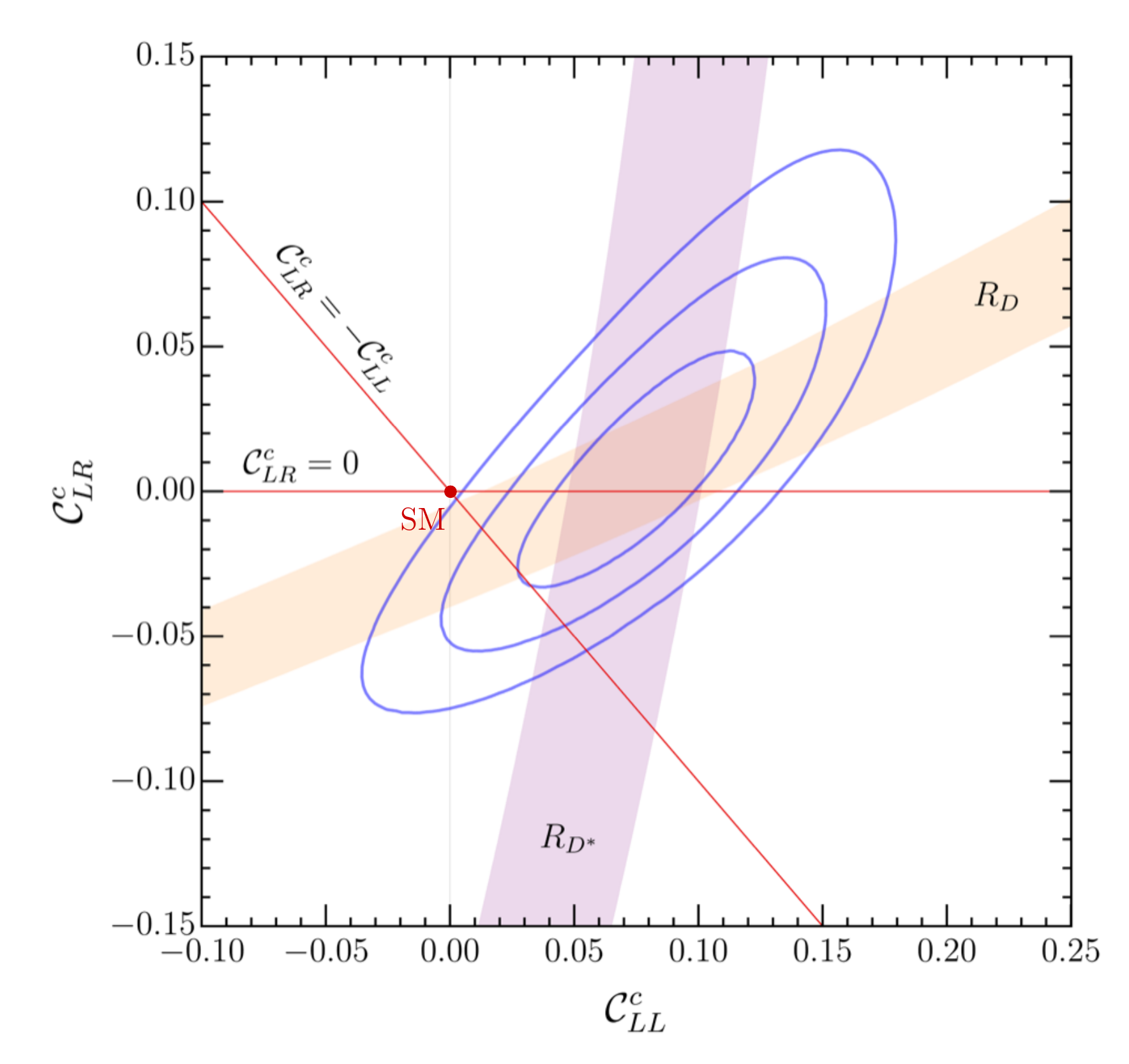}
  \caption{EFT fits to the $b \to s \ell \ell$ anomalies (left) and the $b \to c \ell \nu$ anomalies (right) \protect\cite{Cornella:2021sby}. The coefficients $\Delta C_i$ are the NP contributions ($C_i = C_i^{\mbox{\tiny SM}} + \Delta C_i$).}
  \label{fig:LEFTfits}
\end{figure}

The effective lagrangian for the $b \to s \ell \ell$ anomalies can be written as \cite{Cornella:2021sby}
\be
  \mathcal{L}_{\mbox{\tiny eff}} = -\frac{4G_F}{\sqrt{2}} V_{ts}^* V_{tb} \frac{\alpha}{4\pi} \sum_{\alpha,\ell} C_\alpha^\ell O_\alpha^\ell \,,
\ee
with
\be
O_9^\ell = (\bar s_L \gamma_\mu b_L) (\bar \ell \gamma^\mu \ell) \qquad O_{10}^\ell = (\bar s_L \gamma_\mu b_L) (\bar \ell \gamma^\mu \gamma_5 \ell) \,.
\ee
The result of the fit to the data, using clean observables only, can be seen in Figure \ref{fig:LEFTfits}, where it is worth noticing that the Left-Handed (LH) NP hypothesis (corresponding to $\Delta C_9^\mu = - \Delta C_{10}^\mu$), is preferred over the SM at $5.0\sigma$, including the latest results for $R_{K_S}$ \cite{LHCb:2021lvy}.
Another interesting result can be found in \cite{Isidori:2021vtc}, where the global significance of NP has been estimated taking into account the \textit{look elsewhere effect}, finding a significance of $4.3 \sigma$.
  
For the charged current anomalies, the EFT is \cite{Cornella:2021sby}
\be
\mathcal{L}_{\mbox{eff}} = -\frac{4G_F}{\sqrt{2}} V_{cb} \left[(1 + C_{LL}^c) O_{LL}^c -2 C_{LR}^c O_{LL}^c\right]
\ee
\be
O_{LL}^c = (\bar c_L \gamma_\mu b_L) (\bar \tau_L \gamma^\mu \nu_L) \qquad O_{LR}^c = (\bar c_L b_R) (\bar \tau_R \nu_L) \,,
\ee
and the fit to $R_D$ and $R_{D^*}$ can be seen in Figure \ref{fig:LEFTfits}, resulting in a NP significance of $3.2 \sigma$ over the SM.

\section{A combined explanation?}

It is interesting to investigate the possibility of a combined explanation of the two sets of anomalies.
This seems to be suggested by the fact that both are indications of LFU breaking involving $b$ quarks.
Moreover, at energies above the electroweak scale, NP effects can be parametrised in the so-called SMEFT, i.e. a SM gauge-invariant EFT \cite{Buchmuller:1985jz,Grzadkowski:2010es}.
In this context, the charged current and neutral current transitions involving left-handed fields can be combined into the same operators, since they are related by an $SU(2)_L$ rotation:
\be
(\bar u_L^i \gamma_\mu d_L^j)(\bar \ell_L^\alpha \gamma^\mu \nu_L^\beta) \longleftrightarrow (\bar d_L^i \gamma_\mu d_L^j)(\bar \ell_L^\alpha \gamma^\mu \ell_L^\beta) \,.
\ee
The minimal combined description of the two transitions can be therefore condensed in a single SMEFT operator:
\be
\mathcal{L} = -\frac{2}{v^2} \mathcal{C}_{LL}^{ij\alpha\beta} (\bar q^i_L \gamma_\mu \ell_L^\alpha)(\bar \ell^\beta_L \gamma^\mu q_L^i) \,,
\ee
where a sum over all flavour indices is to be understood.

\begin{figure}
  \centering
  \includegraphics[width=0.4\textwidth]{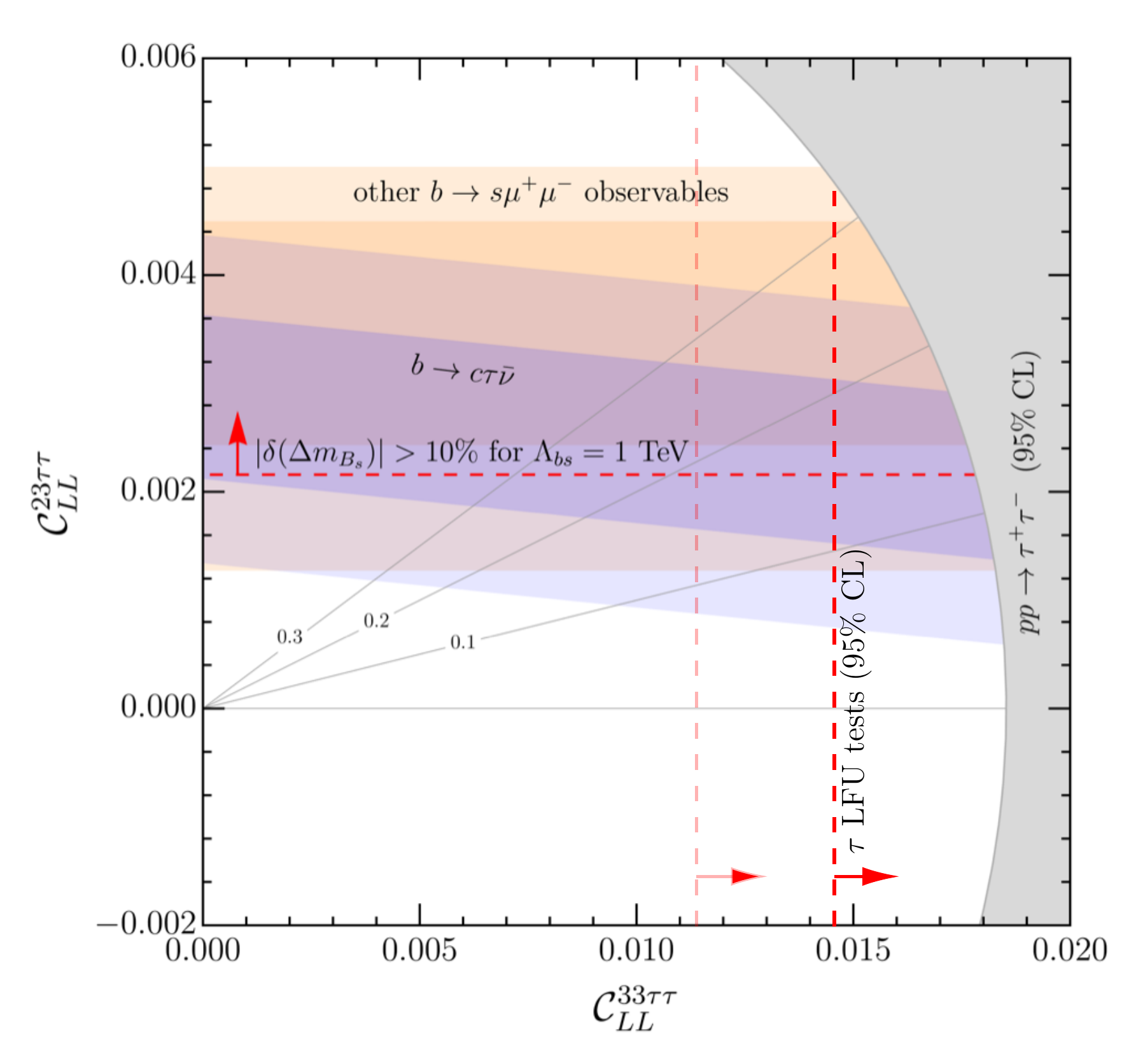}
  \includegraphics[width=0.4\textwidth]{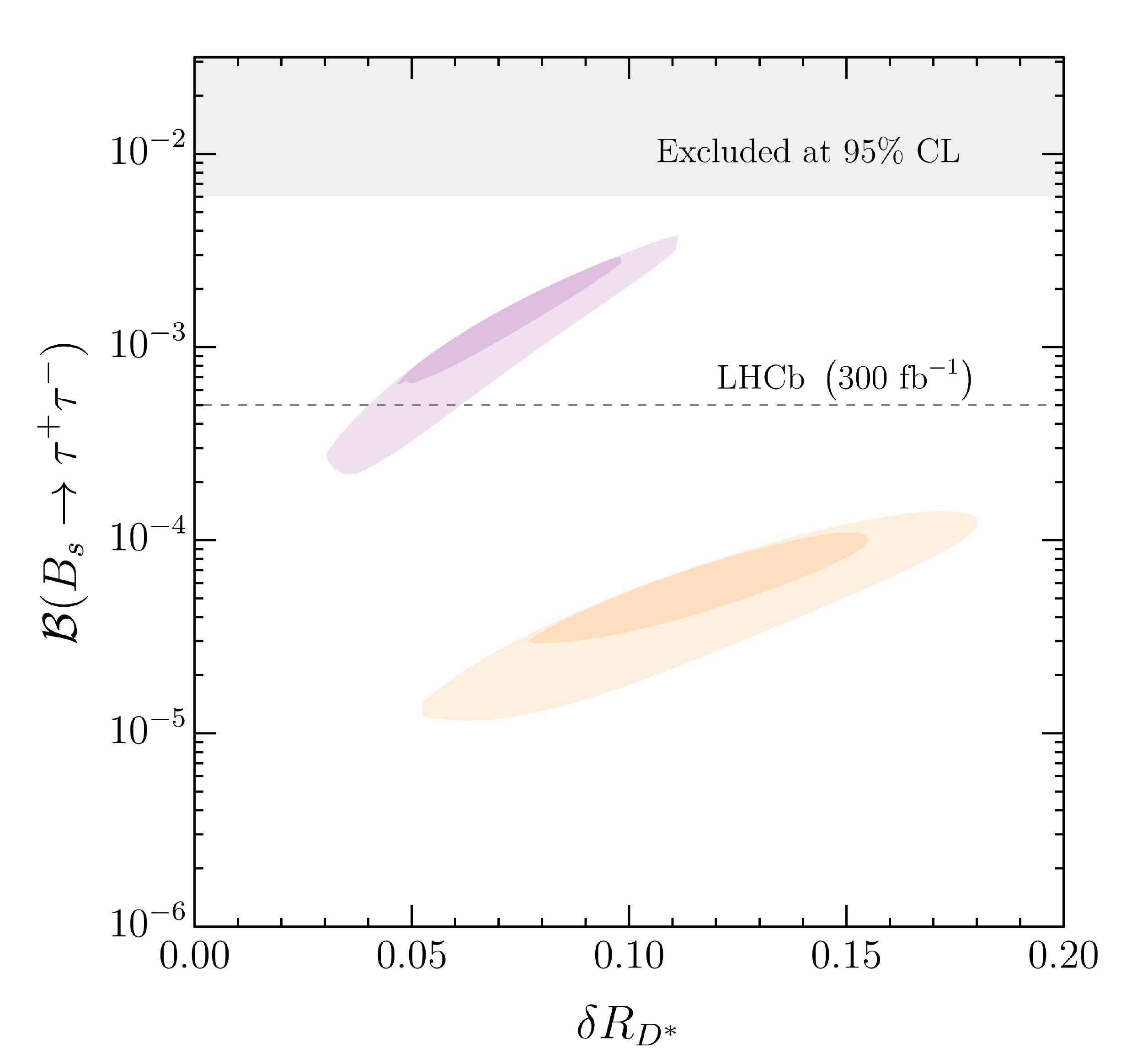}
  \caption{
    \emph{Left panel:}
    Fit of the SMEFT coefficients to the $B$-physics data \protect\cite{Cornella:2021sby}. The clean $b \to s \ell\ell$ observables have been used to fix the coefficient $C_{LL}^{23\mu\mu}$ (not shown in the plot).
    The $b \to c \ell \nu$ anomaly is sensitive to both $C_{LL}^{23\tau\tau}$ and $C_{LL}^{33\tau\tau}$ (down-aligned basis for the quarks), while the other $b \to s \ell \ell$ observables are sensitive to $C_{LL}^{23\tau\tau}$ only via RGE running.
    The vertical dashed lines show the exclusion regions from $\tau$ LFU tests, with (right) or without (left) further assumptions on the UV structure \protect\cite{Allwicher:2021ndi}.
    The horizontal dashed line indicates the bound from $B_s$ mixing, and the grey area is the exclusion region from LHC. 
    \emph{Right panel:}
    Predictions of the $U_1$ model for the $B \to K \tau\tau$ decay, as a function of the size of $R_{D^*}$. In yellow the case of left-handed couplings only, and in purple the case of maximised right-handed couplings, i.e. $|\beta_R^{b\tau}| = |\beta_L^{b\tau}|$.
  }
  \label{fig:SMEFTfit}
\end{figure}

At this point we want to investigate the flavour structure of the possible TeV-scale physics responsible for the anomalies.
The $b \to c \ell \nu$ anomaly, being a deviation from a SM tree-level process, has a much larger overall size than the $b \to s \ell\ell$, which is loop suppressed.
In particular, hypothesizing a new (extremely weak) Fermi-like interaction with semi-leptonic dimension-six operators, the corresponding couplings, in order to explain the effect, need to be \cite{Cornella:2021sby}
\be
G_F^{bc\tau\nu} \sim 10^{-2} G_F \gg G_F^{bs\ell\ell} \sim 4 \times 10^{-5} G_F \,,
\ee
where $G_F$ is the Fermi constant extracted from muon decays. This seems to suggest new physics coupled stronger to third generation fields than with the light generations.
This observation nicely fits into the $U(2)$ paradigm that has been proposed as a way to describe the SM Yukawa couplings \cite{Barbieri:2011ci}. Following the same assumption, one may set $C_{LL}^{33\tau\tau} \sim 0.01$, which is the size needed in order to explain the chraged current anomaly, and scale all other coefficients with powers of small quantities (spurions) $\epsilon_{q,\ell} \sim 0.1$, with one power of suppression for each second generation field, i.e. \cite{Cornella:2021sby}
\be
C_{LL}^{23\tau\tau} \sim \epsilon_q C_{LL}^{33\tau\tau} \qquad C_{LL}^{23\mu\mu} \sim \epsilon_q \epsilon_\ell^2 C_{LL}^{33\tau\tau} \,.
\ee
A two-parameter combined fit to the anomalies can be seen in Figure \ref{fig:SMEFTfit}. In the same plot, the main constraints from other observables are shown, namely:
\begin{itemize}
\item $B_s - \bar B_s$ \textit{mixing}. This constrains the size of the $C_{LL}^{23\tau\tau}$, contributing at one-loop to the meson oscillations.
\item $\tau$ \textit{LFU tests}. Here the $C_{LL}^{33\tau\tau}$ generates, through the top Yukawa, an effective modification of the $W$-couplings to the leptons, yielding a violation of LFU in $\tau$ decays.
\item \textit{high}-$p_T$ \textit{constraints}. These derive from the tails of the $\tau^+\tau^-$ distributions at LHC, and are sensible to both $C_{LL}^{23\tau\tau}$ and $C_{LL}^{33\tau\tau}$.
\end{itemize}

\section{Simplified models: the case of the $U_1$ leptoquark}

When looking for mediators for the semileptonic interactions described above, a good choice is represented by leptoquarks, since they more easily avoid constraints from direct detection and do not give a tree-level contribution to $B_s$ mixing.
Among those, we focus here on the $U_1$ leptoquark, which is a vector transforming as $(\mathbf{3},\mathbf{1},2/3)$ under the SM gauge symmetry. This particular choice has the advantage of being the only single mediator solution to both anomalies, as well as forbidding the $b\to s \nu\nu$ transition at tree level, which would otherwise set tight constraints.
The interaction lagrangian is given by
\be
\mathcal{L}_{U_1} \supset \frac{g_U}{\sqrt{2}} U_1^\mu \left[ \beta_L^{i\alpha}(\bar q_L^i \gamma_\mu \ell_L^\alpha) + \beta_R^{i\alpha}(\bar d_R^i \gamma_\mu e_R^\alpha)  \right] \,,
\ee
where the flavour structure is now contained in the $\beta_L$ and $\beta_R$ matrices. The $U(2)$-like approach can be therefore encoded by having $\beta_{L,R}^{b\tau} \sim 1$ and suppressing the others with powers of a small parameter, e.g. $\beta_L^{s\tau} \sim \beta_L^{b\mu} \sim 0.1$.

Once the size of the parameters has been fixed in order to accomodate the anomalies, one needs to look at the constraints coming from other sectors.
In particular, the striking signatures of the $U_1$ model with the discussed flavour strucure include:
\begin{itemize}
\item \textit{Large enhancement in} $b \to s\tau\tau$ \textit{transitions}, resulting in large (with respect to the SM) branching fractions for the rare decays $B_s \to \tau^+\tau^-$ and $B \to K\tau^+\tau^-$ \cite{Cornella:2021sby} (see Figure \ref{fig:SMEFTfit}).
\item \textit{Lepton Flavour Violating (LFV)} effects in $\tau \to \mu$ transitions, such as $B_s \to \tau\mu$ and $\tau \to \mu\gamma$ \cite{Cornella:2021sby}.
\item \textit{Enhancement of} $B \to K \nu \nu$ \textit{transitions}. Despite being forbidden at tree-level, a sizeable effect is induced at loop level due to the large coupling to $\tau$ leptons \cite{Fuentes-Martin:2020hvc}.
\item \textit{Modification of} $pp \to \tau\tau$ at large $\tau\tau$ invariant mass from $t$-channel $U_1$ exchange \cite{Cornella:2021sby}. 
\end{itemize}

\section{Outlook}

The overall picture emerging from the $B$-anomalies is an interesting one, with evidence for LFU violation in both $b \to s \ell\ell$ and $b \to c \ell\nu$ transitions.
In particular, the significance of the measurements in the $b \to s \ell \ell$ system continues to grow.
The $U_1$ leptoquark, coupled mainly to the third generation, seems to be a promising solution to this puzzle, with the potential to connect the $B$-anomalies with the flavour structure of the SM.
As of today, all low-energy and high-energy data are compatible with this solution, but the allowed parameter space is slowly closing, making sure that in the next few years we should be able to confirm or exclude the model.

\section*{Acknowledgments}

We would like to thank Claudia Cornella, Gino Isidori and Javier M. Lizana for their unvaluable help during the preparation of this talk.
The work of LA is supported by the Swiss National Science Foundation under contract 200021-175940

\end{document}